\title{ICC2017-ERA}

\documentclass[journal]{IEEEtran}
\usepackage{times,amsmath,epsfig,latexsym,amssymb,psfrag,graphicx}
\usepackage{paralist}
\usepackage{booktabs}
\usepackage{color}
\usepackage{setspace}
\usepackage{epstopdf}
\usepackage{subcaption}
\usepackage{mathbbol}
\usepackage[ruled,vlined]{algorithm2e}
\usepackage{citesort}

\IEEEoverridecommandlockouts

\title{ Interference Management for Coexisting Internet of Things Networks over Unlicensed Spectrum}

\author{ Amin Azari and Meysam Masoudi\\
	EECS Scool, KTH Royal Institute of Technology, Sweden\\
	Email: \{aazari and masoudi\}@kth.se}
\IEEEpeerreviewmaketitle

\begin{document}
	\setlength{\textfloatsep}{5pt}
	\maketitle
	\thispagestyle{empty}
	\pagestyle{empty}
	
	\begin{abstract}
		The main building block of Internet of Things (IoT) ecosystem is providing low-cost scalable connectivity for the radio/compute-constrained devices. This connectivity could be realized over the licensed spectrum like Narrowband-IoT (NBIoT) networks, or over the unlicensed spectrum like NBIoT-Unlicensed, SigFox and LoRa networks.  In this paper, performance of IoT communications utilizing the unlicensed band, e.g. the 863-870 MHz in the Europe, in indoor use-cases like smart home, is investigated. More specifically, we focus on two scenarios for channel access management: i) coordinated access, where the activity patterns of gateways and sensors are coordinated with neighbors, and ii) uncoordinated access, in which each gateway and its associated nodes work independently from the neighbor ones. We further investigate a distributed coordination scheme in which, devices learn to coordinate their activity patterns leveraging tools from reinforcement learning.  Closed-form expressions for capacity of the system, in terms of the number of sustained connections per gateway fulfilling a minimum quality of service (QoS) constraint are derived, and are further evaluated using simulations. Furthermore, \textit{delay-reliability} and \textit{inter network interference-intra network collision} performance tradeoffs offered by coordination are figured out. The simulation results highlight the impact of system and traffic parameters on the performance tradeoffs and characterize performance regions in which coordinated scheme outperforms the uncoordinated one, and vice versa. For example, for a packet loss requirement of  $1\%$, the number of connected devices could be doubled by coordination.
	\end{abstract}
	

	\section{Introduction}
	Providing connectivity for billions of smart devices, i.e. Internet of things (IoT), is considered as one major driver of the next generations of wireless networks \cite{5gera,5g_iot,iots,6Gvision,5gIoTsurvey}. While the ecosystem for providing higher data rates in 5G is known, the ecosystem for realization of IoT in 5G and beyond networks is still unclear. The main reason consist in the fact that cost-efficiency plays an important role in the realization of IoT \cite{rep1}. In order to provide a cost-effective smart-home sensor network, the deployment of terminals and access points has to be of \textit{plug and play} character. Also, the sensor's price should be as low as few dollars, which results in simpler terminal devices compared to the existing wireless modules \cite{5gera,cost,5gIoTsurvey,5gIoTsurvey2}.  This simplified device design with limited radio front-end requires a novel system design for IoT because it can create adverse propagation conditions, which increases link failure, retransmission rates, and energy waste, and hence, results in short battery lifetimes. The maintenance cost of smart-home systems will be high if their batteries need frequent replacement due to short battery lifetimes  \cite{emac}. Realizing systems with less human intervention \cite{5gr}, as a big driver of smart-home services are of paramount importance for many applications, e.g. it has been claimed that more than 10 years of battery lifetime is required in some 5G use-cases \cite{5gera}. 
	
	\subsection{Literature Study}
	Early visions on next generation wireless access networks, requirements and enablers, and use-cases including IoT communications could be found in \cite{6Gvision,6gsecurity,i4andhealth}. In \cite{IoTBlockchain},  research gaps in relation with IoT and Industry 4.0 have been investigated, including business models and connection with the emerging topic of block-chain. 
		Comprehensive investigations of IoT communications in the context of 5G, opportunities, recent advances and proposed solutions, and open problems could be found in \cite{5gIoTsurvey,5gIoTsurvey2,jaddehabrisham}.  
	Broadly speaking, there are two major solutions for providing wireless connectivity for IoT devices: solutions over the licensed spectrum, and solutions over the unlicensed spectrum. These two categories are described in the following.
	\subsection*{IoT over the Licensed Spectrum}
	Solutions in the licensed spectrum are mainly updated (upgraded) versions of existing cellular infrastructures which can accommodate IoT traffic on the same radio resources as existing non-IoT traffic is served (on exclusive radio resourced dedicated to IoT) \cite{mag_all}. Examples of updated and upgraded systems are LTE- category M  and NB-IoT, where the latter has been proposed with LTE Rel. 13 in 2015 \cite{ciot,wp}.
	The legacy connectivity procedure in LTE-Advanced (LTE-A) networks includes
	synchronization, connection establishment, authentication, scheduled data transmission, confirming successfulness of data transmission, and connection termination
	\cite{sched}. Random access channel (RACH) of the LTE and LTE-A systems is the typical
	way for IoT devices to establish a connection with the BS, and become connected to
	the network. In RACH, each device randomly chooses a preamble from a set of
	available preambles for contention-based access to the BS. Since the total number of collisions and energy wastages, especially when a massive number of devices try to
	get access to the network \cite{sched}. Once a device successfully passes the RA procedure,
	it sends scheduling requests to the BS through the physical uplink control channel
	(PUCCH), the BS performs the scheduling and sends the scheduling grants back
	through the physical downlink control channel (PDCCH), and the granted devices
	send data over the granted physical uplink shared channel (PUSCH) resources to
	the BS. The connection establishment, scheduling, and scheduled data transmission
	procedures in the existing cellular networks have been designed and optimized for a
	different traffic pattern than the IoT communications. While they work well
	for a limited number of long-length communications sessions, for battery-limited
	IoT devices with a massive number of short-length communications sessions these
	procedures are the bottlenecks. Then, network congestion, including radio network
	congestion and signaling network congestion, is likely to happen \cite{cusm}. Towards isolating the impact of serving IoT communications from non-IoT communications,
	in LTE release 13, use of dedicated resources for IoT communications, i.e. NB-IoT connectivity, has been proposed \cite{ciot}.  NB-IoT technology represents a big step towards accommodation of IoT traffic over cellular networks \cite{nbiot}. Data transmission in NB-IoT takes place in a narrow bandwidth, i.e. 200 kHz, which results in more than 20 dB extra link budget compared to the standard LTE-A systems. While using cellular-based IoT serving solutions provides reliability to the system, these solutions usually have complicated and more expensive radio modules than solutions deployed in the unlicensed band, which also results in shorter battery lifetimes \cite{lif_com,cost}. Then, here we narrow our focus  down to the smart home scenario in which each home is equipped with a private gateway operating in the industrial scientific and medical (ISM) radio band and tens/hundreds of connected smart devices per home, office, or factory.

	\subsection*{IoT over the Unlicensed ISM Spectrum}
	From 1G to 4G, the telecommunications industry has spent a great deal of re-
	sources investigating how to realize high-throughput infrastructure but forgotten
	about scalable low-power low-rate systems in which, energy efficiency and battery
	lifetime are crucially important. Along with telecoms' transition from 3G to 4G
	networks, some 3GPP-independent companies have tried to enter the market and
	provide large-scale IoT connectivity over the unlicensed spectrum, e.g. SigFox and
	LoRaWAN. While reliability over the unlicensed spectrum is a big concern for these
	solutions, they have focused on minimizing the cost per device and maximizing the
	battery lifetime in order to deliver a competitive service at a low cost. Regarding the
	fact that energy consumption in the synchronization, connection establishment, and
	signaling is comparable with, or even higher than, the energy consumption in the
	actual data transmission \cite{nbt}, IoT solutions over the unlicensed spectrum leverage
	grant-free radio access for energy saving \cite{lif_com}. In grant-free access, once a packet
	is triggered at the device, it is transmitted without any handshaking with the BS
	or authentication process. Furthermore, the IoT solutions over the unlicensed spectrum mainly leverage narrowband, or even ultra-narrowband, communications in
	order to cover a large area with the minimum possible energy consumption at the
	device-side \cite{all_comp}. These low-power wide-area (LPWA) IoT networks over the unlicensed spectrum, along with cellular NB-IoT and LTE-M networks, are expected to
	share 60 percent of the IoT market among themselves, a number that is expected to
	grow over time, and hence the competition between LPWA technologies is becoming
	intense \cite{all_comp,ml}. Furthermore, one needs to add Wi-Fi and Bluetooth-powered IoT solutions  to the list of competitors for enabling  IoT connectivity.  SigFox (introduced in 2009) and LoRa (introduced in 2015) are the two most important IoT solutions deployed over unlicensed spectrum \cite{mag_all}. These two solutions along with several other solutions coexist   on the same band, and hence, are vulnerable to collisions when the network scales.  These solutions are referred to as \textit{capillary solutions} because they use a star topology, and hence, data is transferred over installed local gateways by users, where these gateways have Internet connectivity \cite{emac,bid,ist}.  The main design objective in these solutions is to shift complexity from device side to network side, in order to enable connectivity for low complexity devices, which also reduces the price per module significantly \cite{mag_all,all_comp}.

	\subsection{Motivation: Interference in coexisting IoT networks}
	When collecting data from a huge number of devices, the density of the gateways obviously will  be large, at least one per home, and hence, the interference coming from the devices sharing the same resource simultaneously could be significant. According to \cite{mm}, tolerable interference level on each carrier can greatly affect the system performance. Then, regarding  the open-access nature of the ISM band, this impact is even more severe and therefore it requires more investigation. To get insights into the interference problem in the ISM band, one may refer to Fig. 3 of \cite{int1}, in which interference measurements in the European 868 MHz ISM band have been depicted. In this figure, one sees that in use-cases like a business park, the ISM band has been occupied almost all the time. This means that there is a high probability of collision in data transmission over the ISM band in dense use-cases \cite{dis}. Furthermore, regarding the fact in those solutions devices are not easily reachable and controllable through the network, change in communications characteristics of devices, e.g. adapting them to the network traffic, is not possible without human intervention.

	\subsection{Related Works}

	The impact of interference on IoT communications has been investigated in several experiments through radio measurements, including \cite{expev,fingrd,cox,int1}. 
	Due to the crucial impact of interference on IoT communications, especially as shown in \cite{expev,int1},  different studies have investigated approaches for modeling and mitigation of interference in different use-cases \cite{masoudi2021low,masoudi2018grant}. In \cite{eecs}, a cooperative spectrum sensing approach for interference-aware communications has been developed. Autonomous interference mapping for industrial IoT Networks over unlicensed bands has been proposed in \cite{grimaldi2020autonomous}.  Machine learning approaches for sensing and modeling of interference in IoT communications has been investigated in \cite{mlmodel}. The performance of Nb-IoT communications over unlicensed spectrum has been investigated \cite{covNB}.   
	
		The authors in \cite{ivanov2020interference,indorif,3dif} have investigated Iot communications in indoor use-cases, with special focus on the impact of interference. More specifically, \cite{ivanov2020interference,3dif} leverage 	3D Mapping in Dense Indoor IoT Scenarios for characterization of interference and throughput. In \cite{indorif}, decentralized machine learning approaches are used for mitigation of traffic in indoor IoT communications.  One observes that there is a lack of research on interference modeling and management from co-existing IoT networks in indoor use-cases, which is investigated in this work. 
	
	\subsection{Contributions and Structure}
	Here, we  aim at investigating the  impact of (i) system parameters, including communication protocol and propagation environment parameters, as  well as (ii) traffic parameters, including a plurality of devices per square area and  activity profiles of them on \textit{capacity}, \textit{delay}, and  \textit{battery lifetime} as three  key performance indicators (KPIs) of IoT ecosystem.  Then, we present a learning-powered coordination solution for enhancing the KPIs of interest through coordination of activity pasterns of neighboring networks. The feasibility of the proposed solution has been investigated in a smart home scenario, where the results show the operation regions in which, coordinated and uncoordinated  access perform favorably. The major contributions of this work are as follows.
	\begin{itemize}
		\item
		Present an interference management protocol for coexisting grant-free radio access IoT networks by  coordinating activity patterns of devices. 
		\item
		Derive analytical  expressions for the KPIs of interest, including battery lifetime, experienced delay, and outage probability.
		\item
		Present existence of a switchover operation point where beyond this point, the uncoordinated protocol outperforms coordinated protocol in capacity and battery lifetime.
		\item
		Shed light on the increase in the capacity of the system, in handling simultaneous connections, by proper coordination through simulations results. For example, in the smart-home use case,  for packet loss ratio of $1\%$, results confirm more than 100\% increase in the number of sustained connections  per apartment. 
	\end{itemize}
	
	The remainder of this paper is organized as follows. The system model, performance indicators of interest, and key assumptions are introduced in the next section. Analytical analysis and derivation of closed-form analytical expressions for performance indicators are given in section III. Simulation results and discussions  are presented in section IV. In section V, the concluding remarks are presented.

	\section{System Model, Assumptions, and KPIs }
	We consider a smart home network in a building with multiple apartments. One must note that while the analytical expressions in the following are presented in the context of the smart home, they could be easily extended to any other IoT use case, e.g. smart office, factory, business park, and even outdoor IoT applications. Each apartment is equipped with a gateway and multiple sensors distributed randomly in each apartment. The gateway and sensors in each apartment can receive interference from neighboring sensors and gateways. Each sensor sends its measurement data every  $T_r$ seconds. Associated nodes to each gateway use a grant-free access protocol to communicate with the  gateway in a channel with a bandwidth of $W$. Both actuators and sensors have limited battery capacity. Moreover, gateways are connected to the electricity grid and can exchange information with each other using their backhaul already deployed in the building. 
	Let us assume $M$  sensors, $N$ actuators, and one gateway have been deployed in each apartment, where the gateway has been located at the center. The transmit powers of sensors, actuators, and gateway are $P_t^s, P_t^a,$ and $P_t^g$, respectively.  Furthermore, the data rates of sensors and actuators are $R_s, R_a$ respectively.
	Sensors have a reporting period of  $T_r$, and hence, each sensor sends a chunk of data with a length of $D_u$ bits plus $D_o$ bits overhead information to the gateway each $T_r$ seconds. If it doesn't receive ACK in $T_{ack}$, it will resend the packet. Actuators have an operation period of $T_a$, and hence, each actuator sends a packet of length $D_a+D_o$ bits to the gateway each $T_a$ seconds to let the application know that it is awake. If it doesn't receive a response in $T_{ack}$, it will resend the packet. Also, the gateway will send the set of new orders to the actuator, and the actuator will execute them if any. Gateway also broadcasts beacons regularly to let nodes know that it is ready and can receive data from devices. 
	\subsection{Realistic Interference Model for Smart Home Applications}
	From \cite{plo}, one may define the received power from a device to a gateway as:
	\begin{equation}P_r(d)=P_t+A_{tr}-L_{tr}-\text{PL}(d),
	\end{equation}
	where $d$ is the communications distance, $P_t$ is the transmit power, $P_r$ the received power, $A_{tr}$ the gain of transmit/receive antennas, $L_{tr}$ the losses in transmitter/receiver, and $\text{PL}$ is the pathloss. From \cite{plo}-\cite{plo1}, the pathloss between transmitter/receiver can be modeled as:
	\begin{align}\text{PL}(d)=20&\log_{10}(f)+10\delta\log_{10}(d)+
	34.4 (dB)+\nonumber\\
	&+K_f L_fI_{f}+I_s (K_{w_e} L_{w_e}+K_{w_i} L_{w_i}),\label{plm}
	\end{align}
	where $f$ is the carrier frequency in MHz (e.g. 868), $\delta$ is the pathloss exponent, $r$ is the communications distance in km, $I_f$ is 1 if transmitter and receiver are in different floors, and $I_s$ is 1 if transmitter and receiver are in the same floor. Also, $K_f$ represents number of floors between them and $L_f$ the attenuation loss of each floor. Furthermore, $K_{w_e}$ and $K_{w_i}$ represent numbers of external and internal walls between them, and $L_{w_e}$ and $L_{w_i}$ the attenuation losses of external and internal walls. In these expressions, the walls separating two neighbor apartments have been denoted by external walls, while the walls inside each apartment have been denoted by internal walls. Let us consider a brick wall of width 10 cm is between a sensor and gateway in apartment of interest, and a wall of width 20 cm is between a sensor in a neighbor apartment and the gateway in the apartment of interest. Then, from \cite{plo1} one sees that the respective pathloss for these two sensors (if they have the same distance to the gateway) will be 4 and 6 dB, respectively, which shows that neighbor sensors can significantly affect the system performance by making interference. The interference at gateway $ g $ is given by:
	\begin{eqnarray}
	I_g = \sum_{x\in \Phi} a_x \text{PL}_{x}^{g} P_{t}
	\end{eqnarray}
	where $ \Phi $ is a set of interfering nodes and $ a_x $ is a binary variable which is equal to $ 1 $ if the node is transmitting and is $ 0 $ otherwise. Also $ \text{PL}_{x}^{g} $ is the pathloss between node $ x $ and gateway $ g $.
	
	\subsection{Key Performance Indicators  of Interest}
	Given a quality of service (QoS) metric for communications, e.g. number of dropped packets in a period of time, maximum number of supported devices in the system, i.e. system capacity, is the main KPI to be investigated. This KPI indicates whether the set of provisioned radio frequency resources and its respective MAC protocol can handle the required amount of traffic load with a predetermined reliability level or not. Furthermore, battery lifetime is of great interest in IoT applications. This is due to the fact that the maintenance cost of IoT systems will be high if their batteries need to be replaced frequently. Furthermore, a big driver of IoT is realizing systems with less human intervention \cite{emac}, which is in contrast with the need for battery replacement. Then, here we will investigate the impact of protocol, environment, and traffic parameters on the battery lifetime. 
	
	\section{Radio Resource  Management}
	Grant-free access, in which each device sends data to the gateway without the need for prior handshaking, is the underlying multiple access over the ISM band. From the network side, the only restriction comes for the fair use of the  channel, i.e. there are regulations for maximum transmit power and/or  duty cycle of each sub-band.  Here, we go further and investigate how devices, including gateways and sensors, can utilize the channel in a coordinated manner, while there is still no need for resource reservation and signaling. 
	
	\subsection{Coordination for Radio Access Control}
	In order to exemplify the coordination, let us assume the time is divided into communications frames, each containing $K$ subframes. The start of a communication frame could be announced by the gateway through beacons.  Now, we define two operation modes for IoT networks in each apartment: coordinated and uncoordinated access. In the uncoordinated mode, the IoT network in each apartment has full access to the radio resources at all the subframes. However, in the coordinated mode, the IoT network in each apartment has access to the radio resources in dedicated time slots, where the choice of subframes could be done in a centralized or distributed way, as follows.
	
	\subsubsection{Centralized Coordinated Radio Access}
\begin{figure}[t!]
		\centering
		\includegraphics[ width=3.5in]{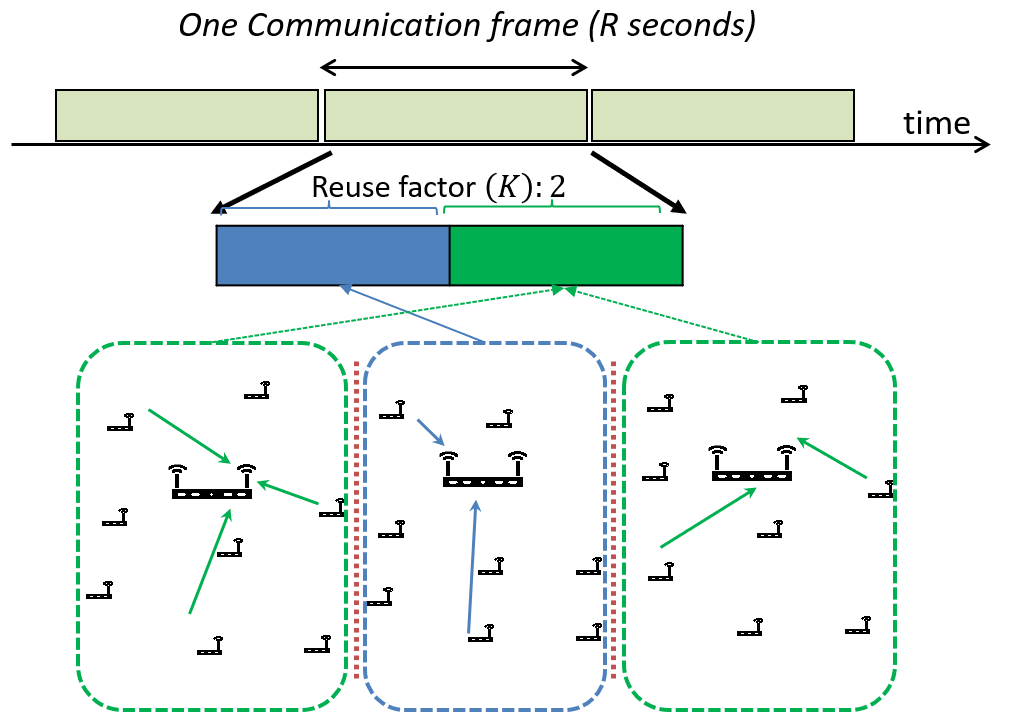}
		\caption{Coordinated access for $\mathcal K$=2, i.e. time frame is divided to two slot, in a centralized way. Orthogonal time slots (blue and green) are allocated to neighbor networks, i.e. the three gateways. The network placed in the middle performs its communications in the first slot of each frame, and the two networks which are on the left and right sides perform their communications in the second slot of each communication frame. }
		\label{cr}
	\end{figure}

	In the coordinated radio access mode, each gateway has an activity pattern, which is coordinated with the neighbors, as depicted in Fig. \ref{cr}. In a beacon-based communications setup, each gateway may send beacons in the allocated time slot for its activity, and its associated devices send packets in response.  Meantime, the gateway broadcasts acknowledgment to the nodes from which data has been received successfully. Even the coordination could be implemented within devices connected to a gateway, i.e., devices could be categorized into $K$ classes, where each class can only send data to the BS in some predefined subframes.
	
	\subsubsection{Distributed Coordinated Radio Access}
	In the distributed coordinated radio access mode, the selection of subframes on which each gateway or device is active is carried out by learning from past communications. Towards this end, each device keeps an index for each action, where the action is defined as the choice of a subframe for transmission and keeps track of successes and failures in transmission over different subframes (actions). Multi-arm bandit (MAB) learning, a low-complexity class of reinforcement learning,   has been recently investigated in literature for adapting communication parameters of IoT devices \cite{mabm,mab7}, and has shown its merits due to its low complexity and superior performance.  The good thing with learning-powered coordination is its ability to adapt itself to the environment in case of change in the environment, e.g. addition of a new gateway. In the following, we present how MAB learning works and 
	could be leveraged in our problem. 
	
	\subsubsection*{MAB Learning for Distributed Coordinated Access}
	
	In reinforcement learning learning, each  device aims at maximizing its objective function, e.g. reliability, battery lifetime, or experienced delay by choosing the best  action, given the rewards of its previous actions (data transmissions).  After choosing the  action at time $t$, the device receives an acknowledgment(ACK) or non-acknowledgment (NACK) as a reward, which is accumulated under the index of each action by 1 and 0, respectively. Furthermore, the device keeps track of the number of times it has visited each action to be able to get the average value of each action. Furthermore, having the number of times an action has been experienced is also useful for determining  the value of each action for re-experience, i.e. exploration, instead of always working based on the best results so far (exploitation). This type of reinforcement learning is commonly described as {\textit{ multi-arm bandit}} in the machine learning literature \cite{sadegh}. Due to its widespread applications in gambling, robotics, etc. MAB learning has been well investigated in the literature, and efficient solutions have been proposed to minimize agent's regret. The interested reader is referred to \cite{sadegh} for further information on MAB problems. In the following, we present an algorithm for distributed coordination of IoT devices in a distributed way using MAB learning.
	
	\subsubsection*{The Algorithm}
	Due to the random nature of interference coming from a wide set of coexisting  IoT devices, we model our problem to a stochastic MAB problem. For stochastic  MAB, the MAB in which each arm's reward is drawn from a probability density function, the upper confidence bound (UCB) algorithm performs close to optimally \cite{sadegh}. Among UCB algorithms, we choose the $\text{UCB}_1$ algorithm \cite{aur}. This is because this algorithm  attains a regret growing at $O(\log n)$, where $n$ is the number of rounds \cite{stosdve}. In this algorithm, the device's aim is to maximize its self-accumulative return (summation of discounted rewards) in the long term. Towards this end, the device needs to follow an optimized tradeoff between  exploration and exploitation, where the former indicates decision epochs in which agent tries different actions even if their previously observed rewards are less than the others, and the latter indicates decision epochs at which agent acts greedy based on the previous rewards. This could be achieved by exploring frequently after changes in the network for becoming up to date and then reducing the exploration rate to zero in order to exploit from the knowledge gained in interaction with the environment. This procedure has been summarized in Algorithm \ref{uuc}, in which $\mathbb A=\{1,\cdots,K\}$, 
	\begin{itemize}
		\item
		$k\in \{1,\cdots,K\}$ represent the indexes of actions,
		\item
		$V_k(t)$ represents the value of action $k$ at time $t$,
		\item
		$A(t)\in \{1,\cdots,K\}$ represents the selected action at $t$,
		\item
		$T_k(t)$ represents the number of times action $k$ has been selected until time $t$,
		\item
		$\xi(t)\in\{0,1\}$ represents the received reward,
		\item
		$Z_k(t)$ represents the accumulated reward for action $k$ until time $t$,
		\item
		$\alpha\in(0,1)$ tunes the exploration/exploitation tradeoff.
	\end{itemize}
	
	\begin{algorithm}[t!]
		\nl Initialization: $Z_k(1)\text{=}0, T_k(1)\text{=}1, \forall k\in \mathbb A $\; 
		\nl \For{$t=1,2,\cdots$}{
			- Update value: $V_k(t)=  Z_k(t)+\sqrt{ {\alpha\log(t)}/{T_k(t)}}$\;
			- Take action: $\arg \max_{k\in \mathbb A}\hspace{1mm} V_k(t) \to A(t)$\;
			- Receive reward: $\xi(t)\in\{0,1\}$\; 
			- Update reward:    $Z_k(t\text{+}1)\text{=}Z_{k}(t), \forall k\in \mathbb A$$\setminus$$A(t)$\;
			\hspace{24mm}   $Z_{A(t)}(t\text{+}1)\text{=}Z_{A(t)}(t)\text{+}\xi(t)$\;
			- Update counter: $T_{A(t)}(t\text{+}1)\text{=}T_{A(t)}(t)\text{+}1$\;
			\hspace{25mm} $T_{k}(t\text{+}1)\text{=}T_{k}(t),\forall j\in \mathbb A$$\setminus$$A(t)$\;
			- \Return $A(t)$\;
		}
		\caption{MAB learning for distributed coordination of co-existing IoT devices}\label{uuc}
	\end{algorithm}
	
	\subsubsection{Uncoordinated Grant-free Radio Access}
	In the uncoordinated radio access mode,  there will not be any coordination among devices in different apartments in connecting to the same gateway, as well as among activity patterns of gateways in different apartments. Each gateway is always on, sends beacons regularly (several times per reporting period of devices), and sends ACKs back per received packet. We consider  ALOHA  to be used for modeling communications between devices and the access points in this scenario \cite{aloha}. 
	
	\subsection{Performance Tradeoffs offered by Coordination}
	Coordination offers two tradeoffs. The first tradeoff consists in transmission delay, which is traded to achieve reliability. In other words, in coordinated radio access, a device waits for a proper time slot dedicated to its network to send data. The second tradeoff is between interference level from neighbor networks and probability of collision with packets from the same network. This is due to the fact that coordinated MAC benefits from reducing interference from devices deployed in neighbor networks at the cost of reducing the time at which devices located in an apartment must send their packets. In section IV, we will see the traffic regions at which coordinated MAC outperforms the uncoordinated MAC and vise versa.

	\begin{figure}
		\centering
		\includegraphics[width=3.5in]{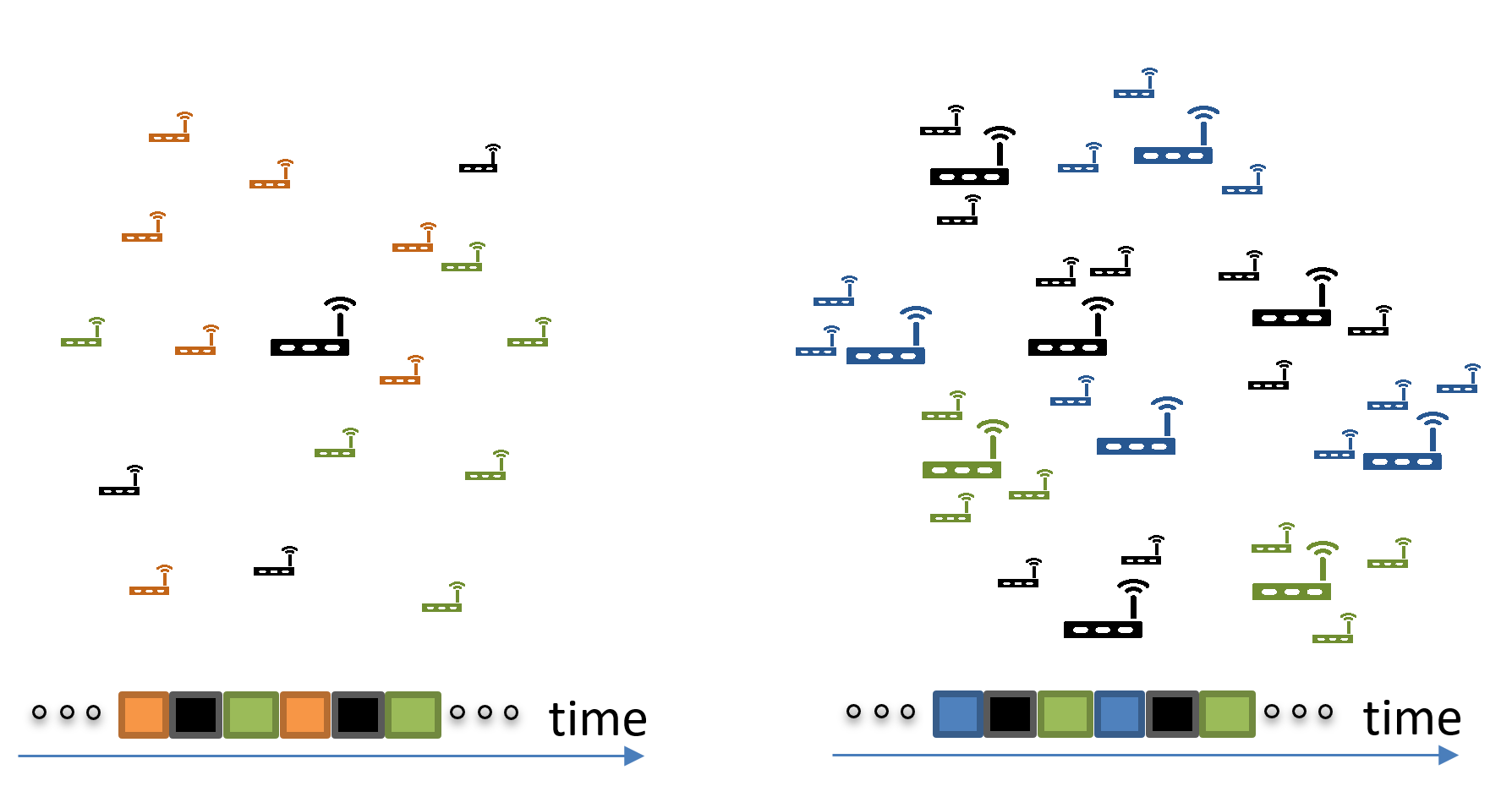}
		\caption{(Left) Coordinated activity management for coexisting IoT devices within a network ($\mathcal K$=3), where devices have learned to transmit in one of the three time periods. For example, the green devices send their packets int he third slot of each frame.  (Right) Coordinated activity management for coexisting gateways, where gateways configure their activities in a distributed way.  For example, the communications between blue gateway and its devices occur in the first slot of each frame.}
		\label{mc}
	\end{figure}
	
	\begin{figure}
		\centering
		\includegraphics[width=3.4in]{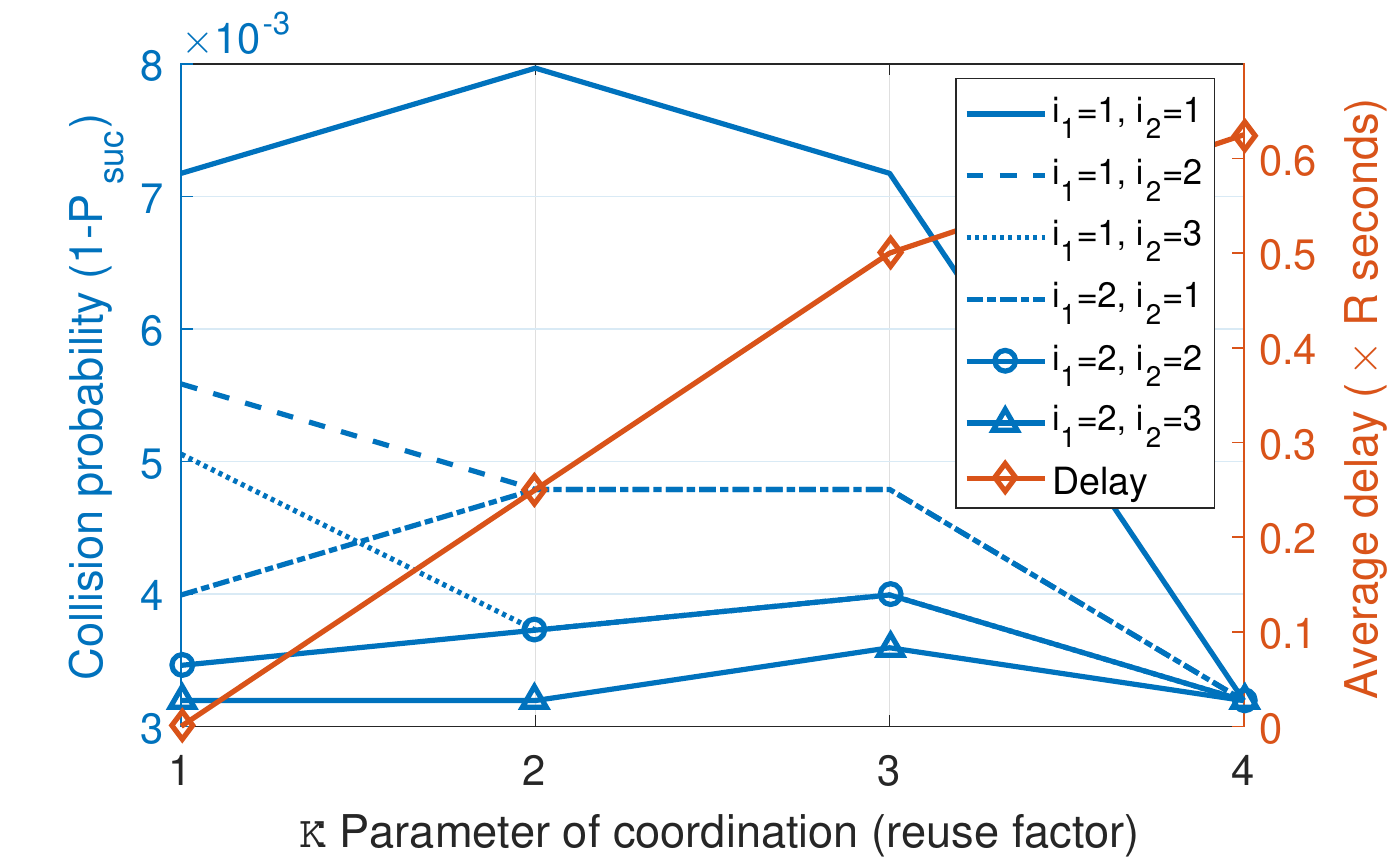}
		\caption{The reliability-delay tradeoff offered by coordination. }
		\label{trf}
	\end{figure}

	\section{ Analytical KPI modeling}\label{anal}
	In this section, we derive analytical expressions for system capacity and expected battery lifetime of connected devices as a function of the system, MAC, and traffic parameters. These expressions enable us to see how different parameters impact the performance tradeoffs.
	\subsection{System capacity}
	We assume packet arrival at each node is a  Poisson process with rate $\lambda$. In case a collision occurs, each colliding node retries at a later time, where the trial time instances are Poisson distributed points with the rate $\vartheta$. We aim to  investigate the maximum number of devices that can be supported in the system while the QoS constraint is satisfied. 
	We start with the analysis of the fully-coordinated MAC, in which there is no interfering neighbor  network in tier 1. Denote by $M$ the number of devices deployed in each apartment that are reusing the radio resources. Then, the  average success probability  is defined as:
	\begin{align}
	P_{\text{suc}}&={\bf E}_{T}\{e^{-2T(M - 1)\mathcal K\max(\lambda,\vartheta)}\},
	\label{es1}
	\end{align}
	in which we have assumed that collision with intra-network's transmissions results in a collision.
	Using Jensen's inequality, \eqref{es1} can be lower bounded as:
	\begin{align}
	P_{\text{suc}}&\approx e^{-2\bar T\mathcal K(M - 1)\max(\lambda,\vartheta)}.\label{es}
	\end{align}
	In the above expressions, $1/\mathcal K$ represents the share of each network from the radio resource, i.e. $\mathcal K$ is the radio resources reuse factor, $T$ is a random variable showing the packet length in the time domain with a mean of $\bar T$, and $\bf E_x$ represents expectation over $x$.  For coordinated access in which there are still $\mathcal N$ interfering neighbors, the successful transmission probability is approximated as:
	\begin{align}
	P_{\text{suc}}&\approx e^{-2\bar T\mathcal  K(M-1+\mathcal N M/i)\max(\lambda,\vartheta)},\label{esc}
	\end{align}
	where $i$ denotes the number of concurrent transmissions from a neighbor network that causes collisions in the gateway of interest. It is clear that $i$ depends on the tolerable interference level of the gateway and the number and material of intermediate walls.  One can extend the above expressions to the case where there are multiple classes of neighbors with different $i$:s, i.e. with different levels of isolations between them, as shown in Fig. \ref{trf}.  
	Finally, one can investigate the performance of uncoordinated MAC by setting  $\mathcal K$ in \eqref{esc}  to 1. 
	
	The success probability affects other KPIs like delay, outage, experienced energy in communications, and battery lifetime. The impact of $P_{\text{suc}}$ on two later KPIs  will be investigated in the next subsection. The experienced delay in data communications can be modeled as a function of $P_{\text{suc}}$ as:
	\begin{align}D_{exp}=\sum_{k=0}^{K_{max}}[k(D_u+D_{bo})+D_u]P_{\text{suc}}(1-P_{\text{suc}})^{k}\label{ds},\end{align}
	where $K_{max}$ is the maximum number of transmissions, $D_{bo}$ is the time in the backoff mode after collision, and $D_u$ is the time spent in a single transmission and listening to receive ACK/NACK afterwards.
	
	Also, considering that a packet is in outage after $\mathcal K$ (re)transmissions (e.g. due to expiration of data or limited number of retransmissions), the outage probability is derived as: 
	\begin{align}P_{out}=1-\sum\nolimits_{k=0}^{\mathcal K-1} P_{\text{suc}}(1-P_{\text{suc}})^{k}.\end{align}
	One sees that  given a QoS constraint on delay performance/outage probability (as well as battery lifetime as we will see in the next subsection), we can find  the minimum $P_{\text{suc}}$, that the system can tolerate. Let us denote the required success probability by the IoT application as  $\mathcal P_{\text{suc}}$. Then using \eqref{es}, the capacity constrained to the success probability of $\mathcal P_{\text{suc}}$, is approximated as:
	\begin{align}C_{sys}=1+\frac{\ln(1/\mathcal P_{\text{suc}})}{2\bar T \mathcal K\max(\lambda,\nu)}.\label{Cs} \end{align}

	\begin{figure}
		\centering
		\includegraphics[width=3.5in]{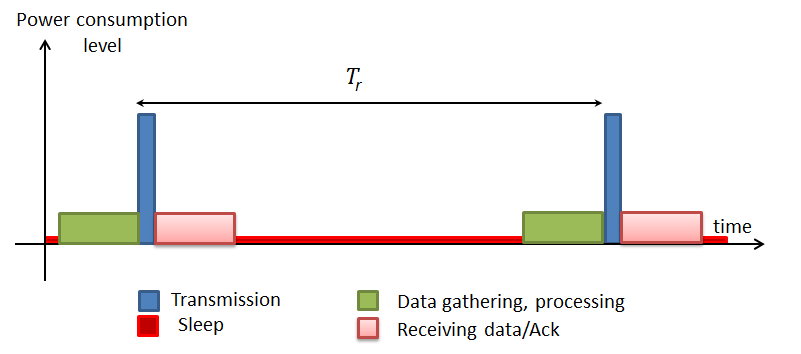}
		\caption{Power consumption of a sensor in one reporting period}
		\label{sen}
	\end{figure}
	\subsection{ Battery lifetime}
	As illustrated in Fig. \ref{sen}, a typical machine node may have different energy consumption levels in different activity modes, including data gathering, processing,
	transmission, and sleeping. For most reporting IoT applications, the packet generation process at each machine device can be modeled as a Poisson process \cite{emac}. Then, the energy consumption of each device can be seen as a semi-regenerative process where the regeneration point is at the end of each successful data transmission epoch. Let us denote the remaining energy of a device as $E_0$, and the average time between two data transmissions as $T_r$.  Now, we define the expected
	lifetime for an IoT  node  at the regeneration point as the product of reporting period
	and the ratio between remaining energy and the average energy consumption per
	reporting period, as follows:
	\begin{align}L_s={E_0}/{E_{\text{PerCycle}}}T_r, \end{align}
	where $E_{\text{PerCycle}}$ is the average consumed energy per cycle. By ignoring energy consumption in the sleep mode, each sensor consumes $E_{sw}$ joules in switching on/off operation, $E_s=P_c T_p$ joules in data gathering/processing (used in circuits, when radio is off), $E_{lis}=P_c T_{lis}$ in initial listening to the  gateway, $1/P_{\text{suc}}E_{dt}$ in data transmission to the gateway, and $1/P_{\text{suc}} P_c T_{ack}$ in receiving acknowledgement from the gateway. In these expressions, $E_{dt}$ is the consumed energy in single-shot data transmission:
	\begin{equation}E_{dt}=(P_c+\alpha P_t^s) (D_u+D_{oh})/R_s,\end{equation}
	$P_c$ is the circuit power, $\alpha$ is the inverse power amplifier efficiency, and$P_{\text{suc}}$ is the probability that data transmission is successful, i.e. the received SINR is greater than the threshold.

	\subsection{Reliability-Delay Tradeoff in Coordination}
	As mentioned in section II.C, coordination offers a reliability-delay tradeoff. In other words, increasing the transmission delay reduces the number of contending nodes in neighbor apartments that are reusing the same resource. Using numerical analysis of \eqref{es}-{ds}, this delay-reliability tradeoff has been depicted in Fig. \ref{trf}.  First-tier interfering neighbors of a square apartment, i.e. 8 apartments consisting of 4 adjacent and 4 diagonal neighbors called group 1 and 2 respectively,  are considered. $i_1$ and $i_2$ represent the minimum number of concurrent transmissions from groups 1 and 2, respectively, which result in collision at the gateway of interest. It is evident that $i_1$ and $i_2$ depend on the tolerable interference level of the gateway and the number and material of intermediate walls. The other parameters used for this analysis can be found in Table 1. In this Fig. \ref{trf}, the $x$-axis represents the reuse factor, i.e. $ \mathcal K$. When $\mathcal K=1$, there is no coordination in operations of gateways. When $K=4$, there are no interfering devices in tier 1. One sees that the experienced delay increases in $\mathcal K$, which is straightforward since nodes wait until the allocated time slots start.  On the other hand, the probability of successful transmission may increase or decrease when $\mathcal K=2,3$. For example, when $i_1=i_2=1$, and  $\mathcal K=2$, the level of increased probability of collision with nodes in the same group increases further than the level of decreased interference from out-of-group neighboring devices. Finally, one sees when $\mathcal K=4$, the probability of unsuccessful transmission has been dropped significantly. This figures out that coordination can improve the reliability of the system at the cost of adding transmission delay, as also shown in Fig 3.

	\section{Performance Analysis}
	In this section, we investigate the impact of system model, MAC, and traffic parameters on the KPIs.   Then, using analytical and simulation results, we will answer the research questions raised in section II. Simulation parameters and values are introduced in Table 1. 
	
	\begin{figure}
		\centering
		\includegraphics[width=3.5in]{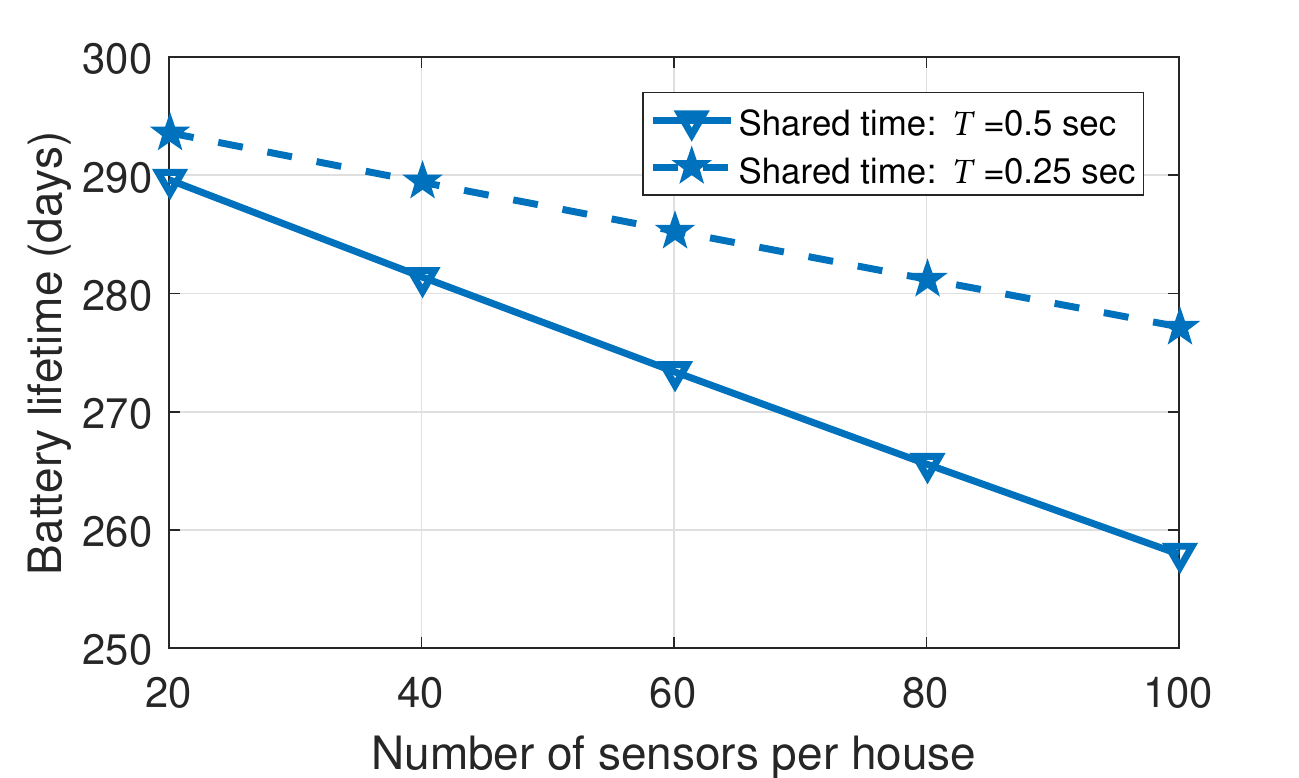}
		\caption{Battery lifetime performance}
		\label{lf}
	\end{figure}

	\subsection{Battery Lifetime Analysis}
	Here, we present analytical results for the battery lifetime analysis. For a typical sensor with an AA battery with capacity of $E_0=1000mAH=3600 J$ and reporting period of 5 minutes, $E_{sw}=1 \text{mJ}$, $P_t^s=10 \text{mW}$, $T_{dt}=(D_u+D_o)/R_s=1 s$, $P_c=1 \text{mW}$, $\alpha=3$, $T_p=5 s$, $T_{lis}=10 s$, $T_{ack}=5s$, and $P_{\text{suc}}=1$, the expected battery lifetime is:
	\begin{align}
	L_s=&\frac{3600\text{ J}}{5\text{ mJ}+1 \text{ mJ}+1/{P_{\text{suc}}\big(36\text{ mJ}\big)}}\frac{5}{24\times 60} \text{  (days)},\nonumber\\
	\to L_s=&295 \text{  (days)}.
	\end{align}
	If $P_{\text{suc}}=0.5$, the resulting battery lifetime will be 160 days. One sees the significant impact of success probability on the battery lifetime. 
	Fig. \ref{lf} represents the expected battery lifetime of sensors versus the number of deployed sensors in each apartment for two cases, i.e.  the packet transmission time, $T$, of  0.25 and 0.5 seconds. As we expect, increasing number of sensors has decreased the success probability. Then, one sees that the battery lifetime also has been degraded by increasing number of deployed sensors. Furthermore, by comparing the two packet lengths in Fig. \ref{lf}, it is evident that by an increase in the traffic load, i.e. when the packet size increases, the impact of number of nodes on the battery lifetime will be critical.

	\begin{table}[t!]
		\caption{Simulation parameters}
		\label{t_sim}
		\begin{tabular}{l|l }
			\textbf{Parameter} &\textbf{Default value}\\
			\hline
			Number of sensors per house&20-100\\
			House radius & $10$ m\\
			Interference threshold&$5\times10^{-9}$ W\\
			Reporting Time&15 min\\
			Packet size &600 bits\\
			Number of neighbors &8\\
			Reuse factor for coordinated access ($\mathcal K$)& 9\\
			Coordination scheme& Centralized\\
			Number of actuators per house&10\\
			Number of gateways per house&1\\
			Data rate of sensors&100 kbps\\
			Loss due to house walls&20 dB\\
			Loss due to room walls&10 dB\\
			Maximum sensor transmit power&10 \text{mW}\\
		\end{tabular}
	\end{table}

	By proper time slot allocation within an apartment, we are able to avoid collision due to the interference. For the dedicated scenario, we have divided the total reporting interval into some subintervals. Each house was given one subinterval to avoid interference from other neighboring houses. It can be seen that when we have less reporting interval, it is more likely to have intra-house collisions, and consequently the packet loss rate increases.  Therefore, on the one hand, we prefer to have more reporting intervals to decrease the rate of collision within a house. While on the other hand, it can cause more collision due to the interference from other houses. Therefore, it is crucial to find the optimum transmission interval value.

\begin{figure}
		\centering
		\includegraphics[trim={0.7cm 0.1cm 1cm 0.5cm},clip,width=3.5in]{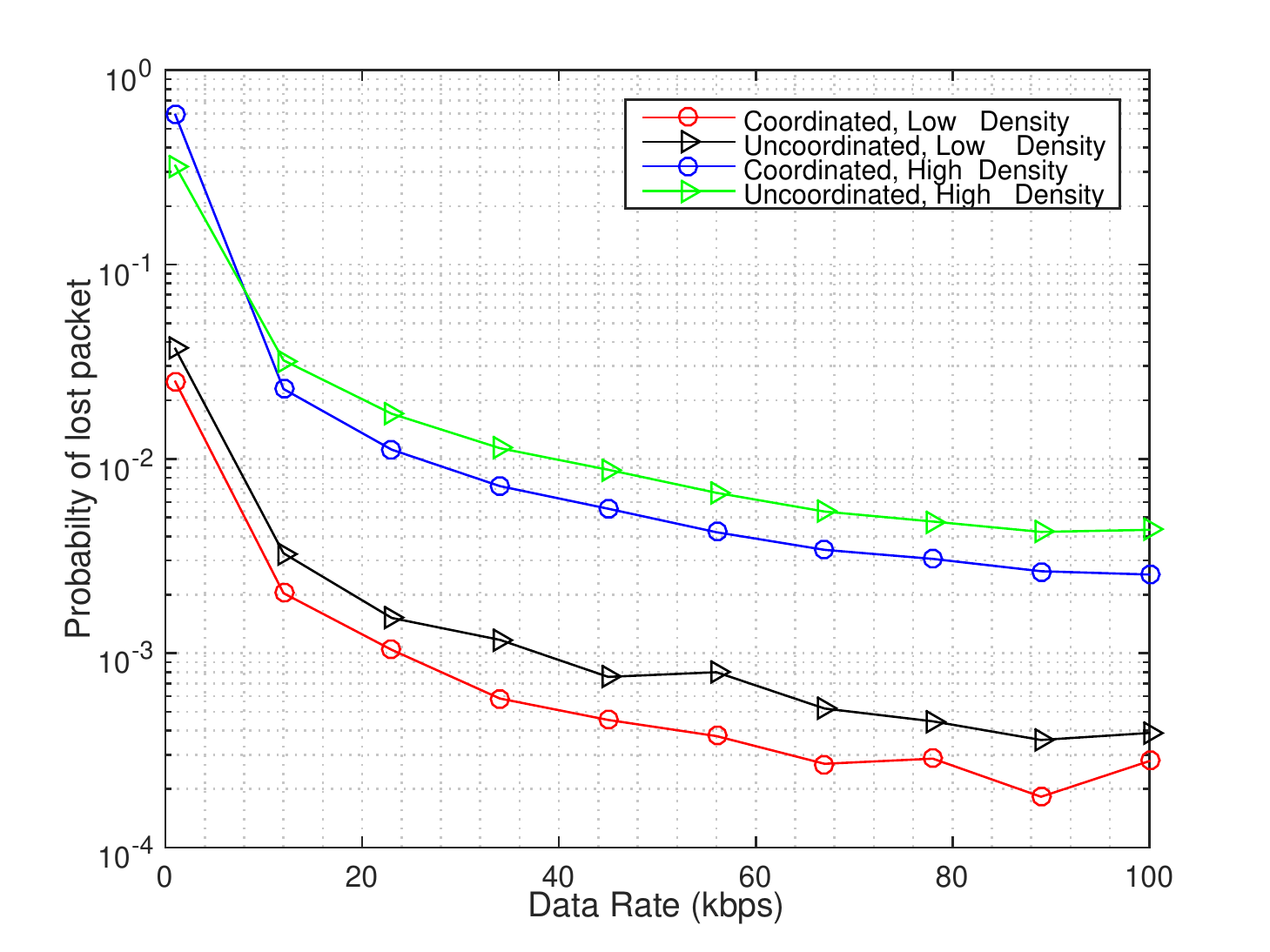}
		\caption{Sensor data rate impact on the system performance}
		\label{DataRate}
	\end{figure}

	In Fig. \ref{DataRate}, we have investigated the impact of sensor data rate on the system performance for the two traffic load regimes, i.e. $ M= 100$ and $M = 10$, respectively, where $M$ represents the number of sensors in each apartment. One must note that due to the fixed length of packets, the higher the transmission data rate is, the shorter packet transmission time will be. Therefore, as one observes from this figure, it is much more likely to receive the packets without any collision with higher data rates. Furthermore, one observes that the merits of coordinated access increase by decreasing the packet transmission time.
	
\begin{figure}
		\centering
		\includegraphics[trim={0.7cm 0.1cm 1cm 0.5cm},clip,width=3.5in]{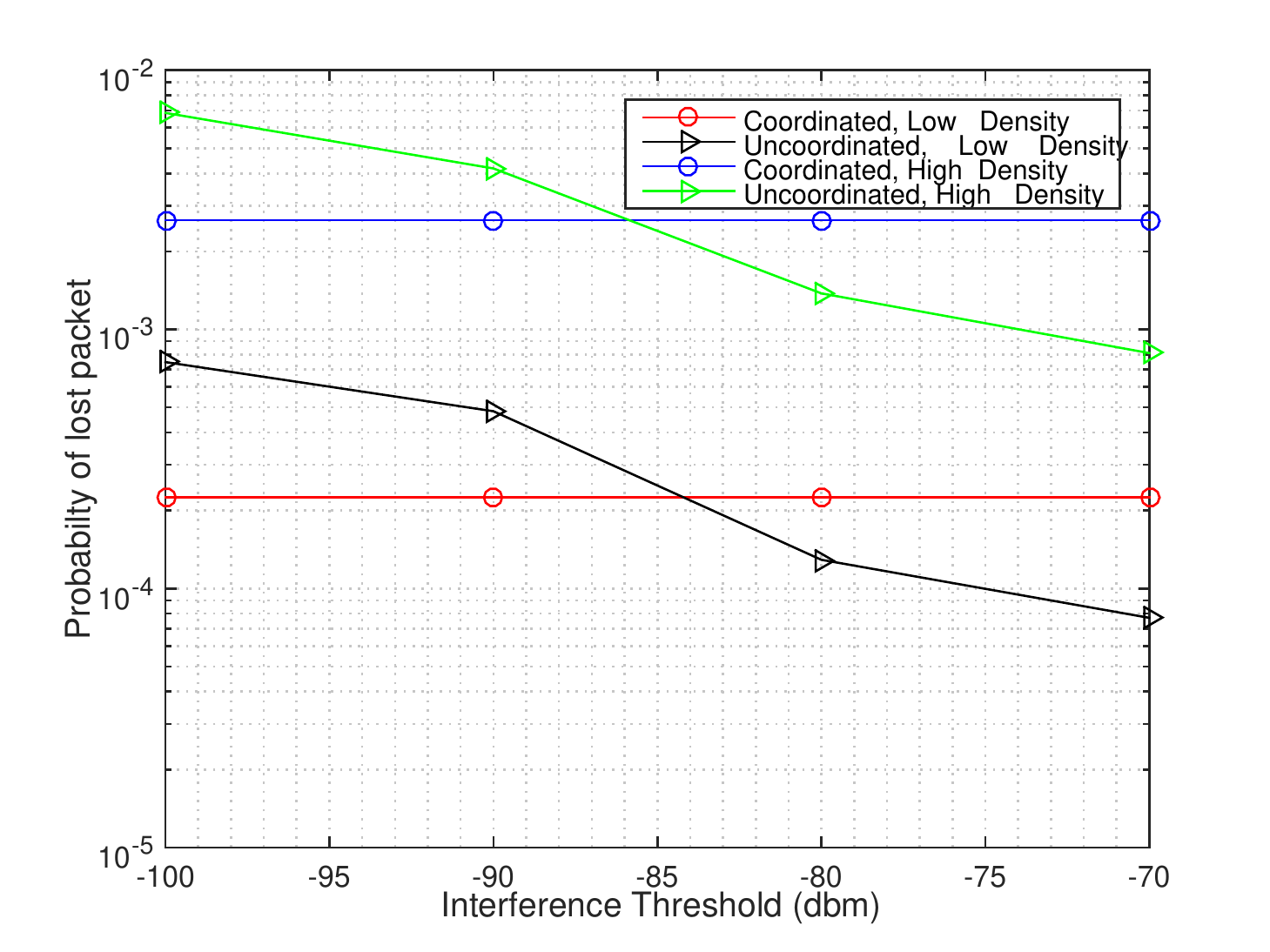}
		\caption{Interference sensitivity of gateways}
		\label{Interference}
	\end{figure}

	In Fig. \ref{Interference}, we have investigated the sensitivity of gateways to the interference. For the coordinated case, gateways do not receive any interference from other apartment sensors and the packet loss is only due to the intra-apartment packet collision, and hence, they have a shorter window time to transmit. In this case, gateways are insensitive to the interference. In contrast, for the uncoordinated case, we have an effective region for the interference threshold in which the performance of the system is controlled by the interference.  For instance, for the high load case, if gateways are highly sensitive to the interference, e.g. very low interference threshold, it is better to apply the coordination scheme. Because almost every interfering packet transmission causes collision while in the coordination case, we do not experience the collision due to the interference. Also, for the interference insensitive case, the collisions are dominated by other parameters such as sensor density rather than interference. The same analysis applies to the low-density case.    
	
	\begin{figure}
		\centering
		\includegraphics[trim={0.7cm 0.1cm 1cm 0.7cm},clip,width=3.5in]{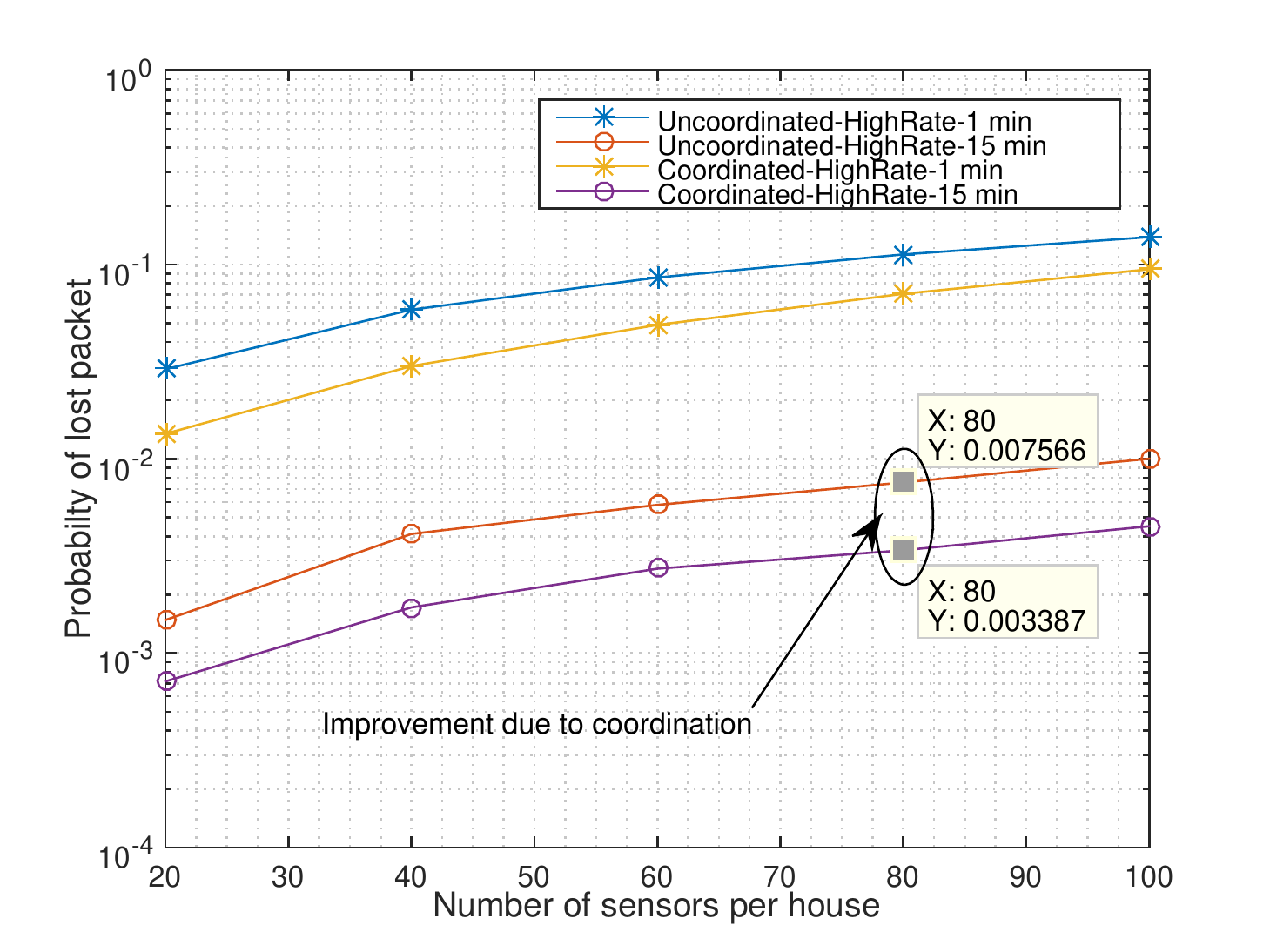}
		\caption{Impact of sensors density  on the system performance. The legend shows the choice of coordination, the rate traffic arrival rate (high and low respective to $M$= 100 and 10 respectively), and the reporting time.}
		\label{SensNum-RepTime}
	\end{figure}

	In  Fig. \ref{SensNum-RepTime}, we have depicted the impact of sensors density on the packet loss ratio. For fixed packet loss ratio, the number of connected devices could be increased by coordination between gateways. e.g. for packet loss ratio $10^{-2}$, the number of connected devices could be doubled (from 40 to about 80). Also, when sensors send data with a low data rate, in the congested case, one can not gain from coordination because the collision is dominated by intra-house collisions. One sees that the expected value from \eqref{Cs} matches well with the result from simulations, and hence, the derived results in section \ref{anal} can figure out system behavior when it scales.
	
	One must note that the scope of this work in performance evaluation has been limited to  the evaluation of performance once the coordination is done, and comparison of it with the case without coordination. On the other hand, the coordination itself could be a subject of further research, as a distributed approach for coordination like MAB could result in a different grouping to another distributed approach without machine learning, and for sure with the centralized approach. Then, a more comprehensive study can shed light on the overall performance, when the coordination and MAC are investigated together.

	\section{Conclusion}
	Providing low-cost scalable connectivity for things in indoor environments is an important enabler of IoT. In this paper, the performance of wireless sensor/actuator deployment in an indoor environment composed of several apartments, each equipped with a single gateway and tens of devices, has been investigated. Centralized coordinated, distributed coordinated by reinforcement learning,  and uncoordinated gateway operation policies have been introduced, and capacity of the system, constrained to a QoS requirement, as well as expected battery lifetime of connected devices have been investigated using closed-form analytical expressions and Matlab simulations. The simulation results confirm the existence of a switchover operation point where beyond that, coordinated protocol outperforms the uncoordinated one. Also, the results confirm that it is possible to increase system capacity by coordination, e.g. for packet loss ratio of $1\%$, the number of connected devices per apartment could be almost doubled, e.g. from 35 it could be increased to 80.

	\ifCLASSOPTIONcaptionsoff
	\newpage
	\fi

	\bibliographystyle{IEEEtran}
	\bibliography{bibl}
	
\end{document}